# A CLASSIFICATION ALGORITHM TO RECOGNIZE FAKE NEWS WEBSITES[*]

Davide Bennato[1], Giuseppe Pernagallo[2] and Benedetto Torrisi[3]

*1 Department of Humanities, University of Catania, (e-mail: dbennato@unict.it)*
*2 Department of Economics and Business, University of Catania, (corresponding author, e-mail: giuseppepernagallo@yahoo.it)*
*3 Department of Economics and Business, University of Catania, (e-mail: btorrisi@unict.it)*

**Abstract**

"Fake news" is information that generally spreads on the web, which only mimics the form of reliable news media content. The phenomenon has assumed uncontrolled proportions in recent years rising the concern of authorities and citizens. In this paper we present a classifier able to distinguish a reliable source from a fake news website. We have prepared a dataset made of 200 fake news websites and 200 reliable websites from all over the world and used as predictors information potentially available on websites, such as the presence of a "contact us" section or a secured connection. The algorithm is based on logistic regression, whereas further analyses were carried out using tetrachoric correlation coefficients for dichotomous variables and chi-square tests. This framework offers a concrete solution to attribute a "reliability score" to news website, defined as the probability that a source is reliable or not, and on this probability a user can decide if the news is worth sharing or not.

---

[*] Please do no quote this work without the authors' permission.

# 1. Introduction

The internet age has redefined the idea of information in all its manifestations: the idea that information can spread almost everywhere in less than seconds is exciting and alarming at the same time. Everyone can propose opinions on every issue creating a chaotic environment where orientation is very often lost. The phenomenon of "fake news" has assumed huge proportions in these years and has showed all the flaws in this system, consequently, the need for truth has grown exponentially not only for internet users but also for authorities and companies. To answer to this necessity, several tools have been developed to detect the veracity of a news such as *Hoaxy* or *Botometer*. In this paper we aim to approach the problem from a higher perspective: instead of classifying a news, we tried to classify its source producing a score on how much reliable the website that originated the news is. In this way, the user will have a numerical datum to decide if the source is trustworthy or not. The algorithm that we propose in this paper is based on logistic regression and assigns a probability that a site is fake based on few predictors obtainable on the website of the originator. The algorithm worked well on our dataset reaching a satisfactory accuracy, offering a concrete solution not only for users but also for owners of platforms, damaged by the diffusion of erroneous information, and for authorities, to reestablish trust in institutions.

# 2. The socioeconomic impact of fake news and the role of the policy maker

Fake news can be defined as unfounded information that mimics the form of reliable news media content. In most cases, originators of fake news lack the structure that characterizes reputable editorial companies, an aspect that is easily recognizable from the website of the source. The phenomenon of fake news can be considered as a form of misinformation or disinformation (Lazer et al., 2018): misinformation occurs when false information is shared without intentionally harm whereas disinformation is false information knowingly shared to mislead people (Wardle and Derakhshan, 2017). Although fake articles are better known, recently, for their political content, they can potentially convey any type of information with equal detrimental effects.

A famous case that illustrates the deep impact of fake news on economic variables was the ImmunoCellular Therapeutics case[2]. ImmunoCellular Therapeutics is a clinical-stage biotechnology company which develops immune-based therapies for the treatment of cancer. On January 18, 2012, an article published on Seeking Alpha reported that the company had discovered an important cancer treatment, cheaper than the existing products. This news, which resulted to be false and architected by the company and the author of the article, pushed up strongly the stock price of the company. Figure 1 shows the evolution of ImmunoCellular Therapeutics stock price before the diffusion of the fake news and after the fake news. The fake news date is indicated by the first dashed line, in red. It is evident that after the spread of this false information the market overreacted increasing the company's stock price. The second dashed line, December 2013, indicates the moment of truth. A discouraging clinical update on the new product of the company caused the price to fell drastically. During 2018 ImmunoCellular Therapeutics stock has been traded at less than $0.50.

---

[2] https://www.ft.com/content/a37e4874-2c2a-11e7-bc4b-5528796fe35c

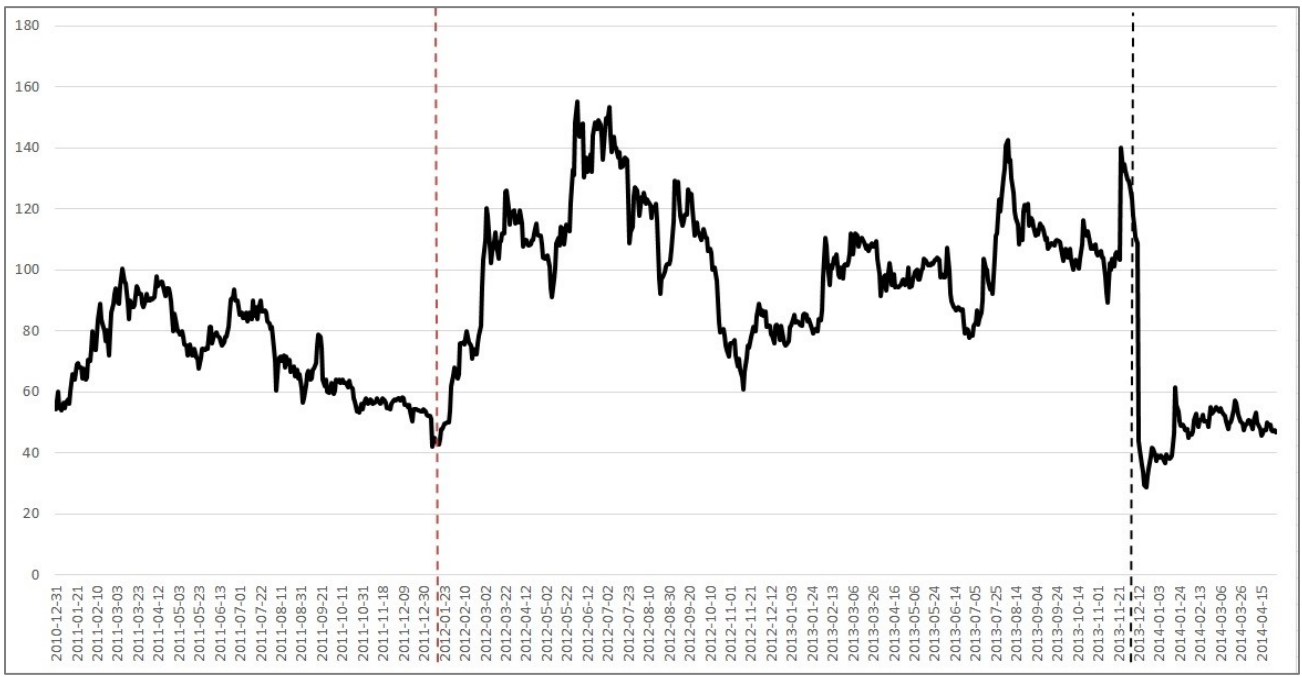

*Figure 1.* ImmunoCellular Therapeutics adjusted stock price from 2010-12-31 to 2014-04-30. Source: our elaboration of yahoo! finance data.

Fake news has also a reputational impact on social networks. Social networks that promote deceiving news lose credibility and rise the hostility of users. For this reason, established social networks invest huge amounts of money to prevent the spread of hoaxes among users.

Even public authorities are concerned about the problem because it undermines the foundations of the democratic information system, furthermore, the diffusion of hoaxes is dangerous on many points of view. From a political perspective, fake news may be employed to drive public opinion towards certain voting choices. From an economic perspective, we have seen in the case of ImmunoCellular Therapeutics how fake news can affect investment or consumption decisions. Finally, from an ethical perspective, governments should always pursue and promote transparency of information for all citizens. The case of Italy is emblematic; as reported by Tambini (2017, p. 13), "*In February 2017 a draft law was introduced to the Italian Parliament in response to the issue of 'Fake News'. This attempted to criminalise the posting or sharing of 'false, exaggerated or tendentious news', imposing fines of up to 5000 Euros on those responsible. In addition, the law proposed imprisonment for the most serious forms of fake news such as those that might incite crime or violence, and also imposed*

*an obligation on social media platforms to monitor their services for such news*". Of course, these measures should find a balance between the need for truthful information and freedom of expression, but this is not the place for such discussions.

**3. The process of diffusion of a fake news**

The causes of propagation of fake news are not different from those that characterize common news. Lee and Ma (2012) designed and administered a survey to 203 students in a large local university and, via Structural Equation Modeling, found out that important determinants of news sharing are:

- gratification of information seeking;
- socializing and status seeking;
- prior experience with social media.

Vosoughi, Roy and Aral (2018) described the process of a rumor cascade on Twitter. The process starts when a user tweets an argument in the form of text, images, videos or links to online articles. Via retweeting, the rumor is propagated to other users based on the dimension of the network in which the originator is placed. Because nowadays social networks are highly interconnected, the process is accelerated by the propagation of the rumor on other platforms. This diffusion process can be characterized by several cascades, each for every user that, independently from the others, originates a tweet regarding the same claim. The dimension of each cascade depends on how many times these tweets are retweeted. Figure 2 shows an example of a diffusion network of a news obtained via *Hoaxy*. The title of the news is "Louis Farrakhan Chants 'Death to America' in Iran", the plot represents the number of tweets that contain this news as an object. We can see that the two principal cascades are the profiles of Brigitte Gabriel and Donald Trump Jr., from their accounts a series of retweets expands densely. We omitted from the plot isolated cascades, i.e. users that independently shared the news that was retweeted only few times. Bigger nodes represent bigger users in term of connections and the color of nodes indicates the similarity of accounts to humans, from light blue (human like) to red (bot like). This network shows how the news spreads in only seven days: a fake news can spread quickly and widely and once it is entered in the system it becomes difficult to stop it. The best way to arrest the propagation of a fake news is to avoid its sharing. Because very often it is difficult to verify the goodness of the content of the news (think of scientific fake news with highly sophisticated terms) a good alternative is to verify the goodness of the original source of the article. Adopting this strategy, a user can have a glimpse on how reliable the news is and may choose to arrest its propagation.

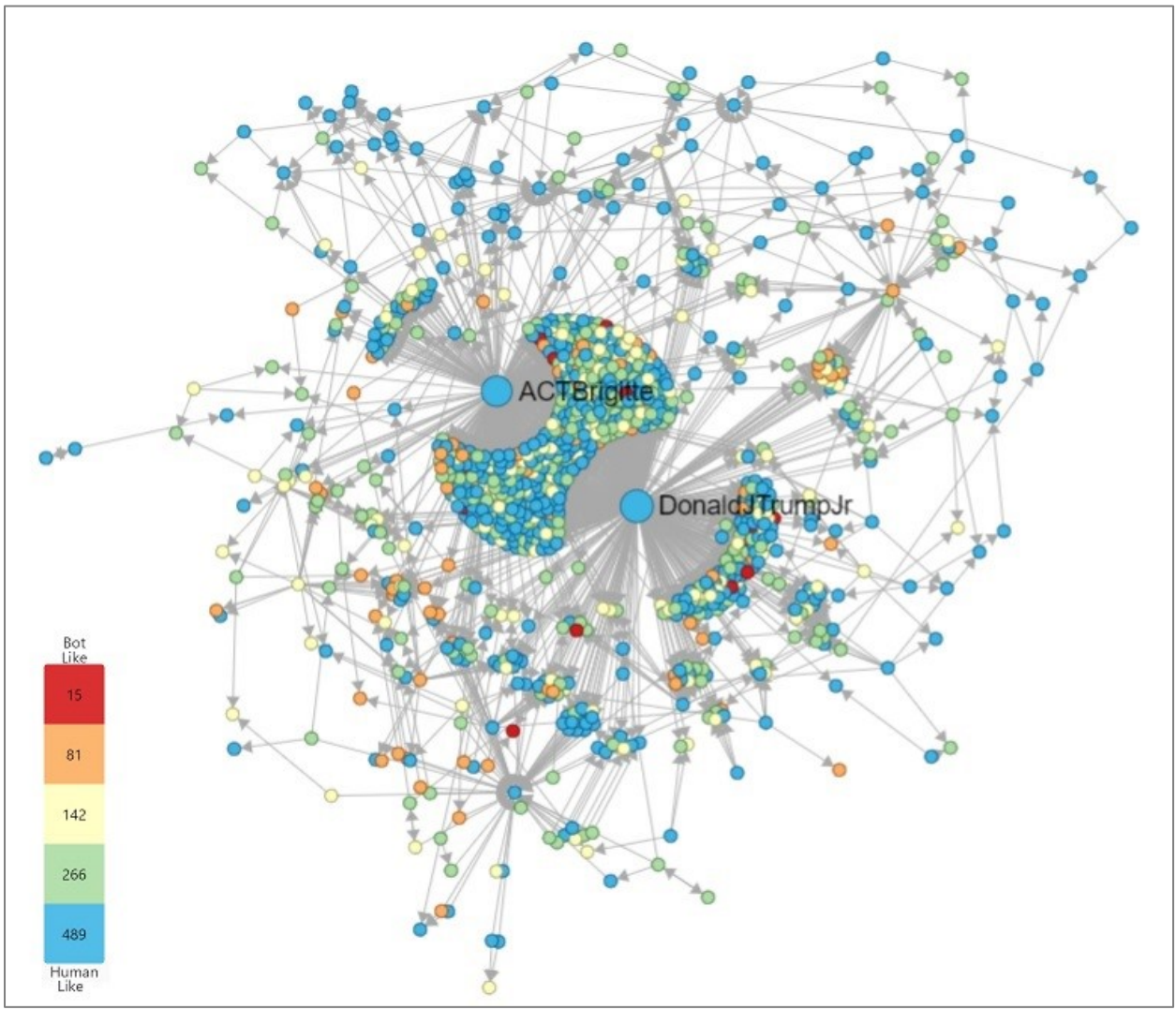

***Figure 2.*** *Diffusion network of the news "Louis Farrakhan Chants 'Death to America' in Iran" from 05/11/2018 to 17/11/2018. Source: elaboration via Hoaxy.*

The problem is that users may have not the skills, the time or the will to distinguish a reliable source from a deceiving source, for this reason we propose in this paper a simple tool to assess the probability that a website is not reliable.

## 4. Methodology and results

As pointed out by Figueira and Oliveira (2017, p. 820), there are two main approaches to face the problem: "*human intervention to verify information veracity*", such as the International Fact Checking Network (IFCN) that allows users to signal fake articles, and "*using algorithms to fight algorithms*". Our attempt falls within the second approach, but differently from already available tools such as *Hoaxy* or *Botometer*, we focused on the source of the news.

The dataset used in this study has been prepared choosing 200 confirmed fake news websites[3] and 200 established news websites from different Countries. We mainly gathered websites from United States, Britain, India, Italy, Philippines, France, Germany, Mexico, Spain and some other Asian sources to dispose of an heterogenous dataset. The choice of 400 websites it is not derived from a specific sample design but was obligated by the lack of data of confirmed fake news websites, furthermore, we thought of it as a pilot dataset to test our algorithm and see how it performs. We chose logistic regression to quantify the reliability of a website because it gives us probability that the dependent variable assumes one of two possible outcomes: the website produces fake news (Fake news website = 1) or the website is reliable (Fake news website = 0). The choice of a logit makes the interpretation of the results easier than other models. For example, the probit model holds similar results but has a more intricate construction. Mathematically the model is:

$$P(\text{Fake news website} = 1 | X_1 = x_1, \ldots, X_5 = x_5) = \frac{\exp(\beta_0 + \beta_1 x_1 + \cdots + \beta_5 x_5)}{1 + \exp(\beta_0 + \beta_1 x_1 + \cdots + \beta_5 x_5)} \quad (1).$$

We used 5 dummy variables to predict the dependent:

$X_1$ = *Padlock*, a dummy variable equal to 1 if the website uses the SSL protocol (a data transfer security standard) or the TLS protocol, 0 otherwise;

$X_2$ = *Contact*, a dummy variable equal to 1 if the website has a "contact us" section or something similar ("connect with us", "gives us a tip", etc.), 0 otherwise;

$X_3$ = *Telephone*, a dummy variable equal to 1 if the website makes available a telephone and/or a fax number, 0 otherwise;

$X_4$ = *About*, a dummy variable equal to 1 if the website has an "about us" section or something similar ("information", "who we are", etc.), 0 otherwise;

$X_5$ = *Terms&Conditions*, a dummy variable equal to 1 if the website has a "terms and conditions" section or something similar ("terms", "legal notes", "terms of use", etc.), 0 otherwise.

The inclusion of these variables is justified by the fact that established news websites clearly expose these elements, which are manifestation of an organized structure compliant with editorial norms and processes. The advantage of this model is that uses only the website to take the needed inputs and compute a probability that the source is reliable.

Figure 3 shows the entire process of recognition. The process starts when a user doubts on the veracity of a news. Inside the circle are reported the operations of the machine. Hypothesize that it is available a tool based on our model, at this point the user insert the URL of the website and the algorithm operates a first screening recurring to an internal database: if the domain or the name of the website explicitly mimics or copies an older and established source of news, the algorithm attributes to that website probability of 1, which means that it is a fake news website. If it is not the case, the algorithm computes the probability using our logit model producing a certain probability $x$. In the last stage of the procedure the user receives this probability and decides, based on her tolerance threshold $T$, to share or not to share the news.

---

[3] There are several articles online that report this information such as the page on Wikipedia, "List of fake news websites", blog.feedspot.com/fake_news_blogs/, usnews.com or the Italian website bufale.net. For other websites it was easier since they mimic existing reputable website domain or deliberately share absurd and ironic news.

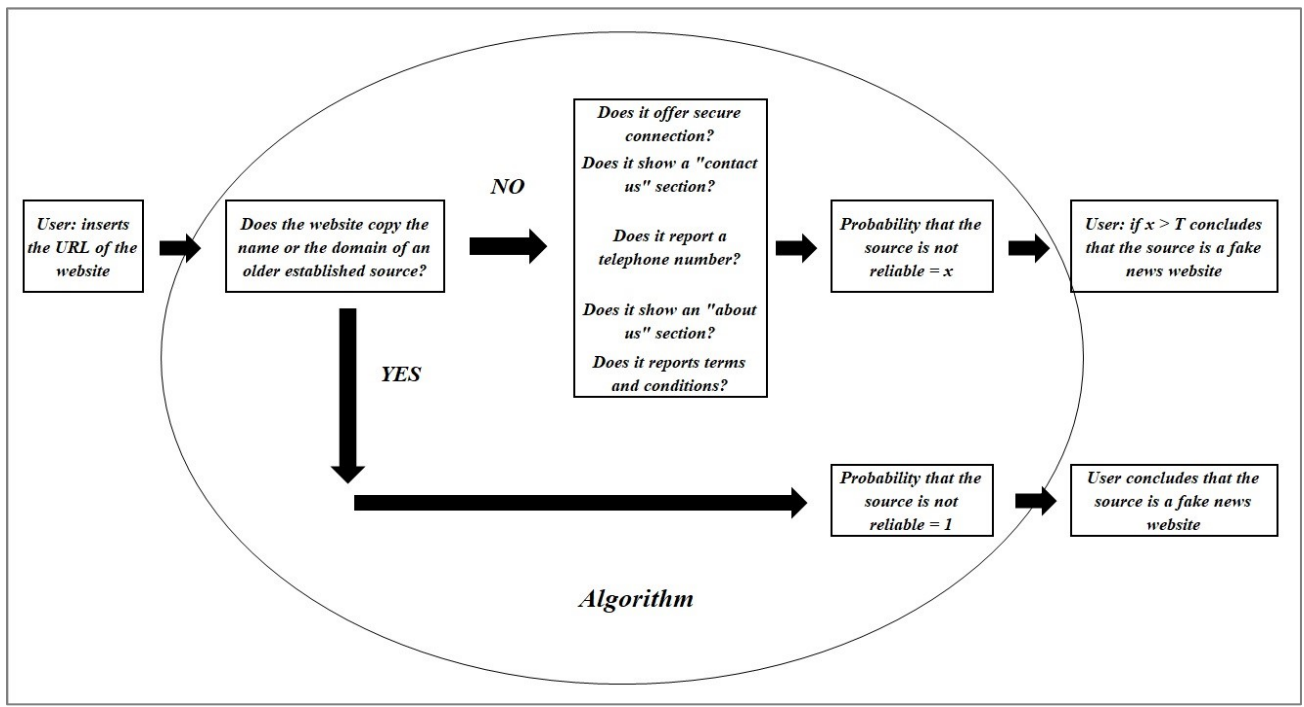

*Figure 3.* *The process of recognition of a fake news website using our model. Source: our elaboration.*

We computed a tetrachoric correlation matrix to test the level of agreement between our dichotomous variables as showed in Table 1. The first column of the matrix shows the agreement between our dependent variable and all the independent variables. There is a strong and negative agreement between the variable Fake news websites and all the considered explicative variables, whereas this agreement is only moderate for the variable About; this means that when one of this section is present the website tends to be more reliable, i.e. the dummy Fake news website tends to assume value 0. The correlation between all the explicative variables is tendentially moderate but highly significant and positive. A positive value means positive agreement, therefore websites that present one of the considered sections tends to present also the other ones.

**Table 1.** Tetrachoric correlation matrix for dichotomous variables. *indicates 0.01% significance.

|  | Fake news website | Padlock | Contact | Telephone | About | Terms & Conditions |
|---|---|---|---|---|---|---|
| *Fake news website* | 1 |  |  |  |  |  |
| *Padlock* | -0.8398* | 1 |  |  |  |  |
| *Contact* | -0.7293* | 0.5321* | 1 |  |  |  |
| *Telephone* | -0.7673* | 0.6689* | 0.8673* | 1 |  |  |
| *About* | -0.3263* | 0.316* | 0.7523* | 0.3285* | 1 |  |
| *Terms&Conditions* | -0.729* | 0.6532* | 0.6899* | 0.503* | 0.4608* | 1 |

At this point we questioned whether the association between the independent variables and the outcome variable is due to chance or statistically significant. We therefore used the chi-square test of independence to test whether the association is significant or not. Table 2 reports the results of the test: it clearly emerges that there is association between all the independent variables and the dependent variable, and this association is highly significant.

**Table 2.** Chi-square test of independence.

| Associated variables | Empirical test statistic | p-value | Expected frequency assumption met? |
|---|---|---|---|
| Fake-Padlock | 146.4103 | <0.0001 | yes |
| Fake-Contact | 74.3226 | <0.0001 | yes |
| Fake-Telephone | 114.1029 | <0.0001 | yes |
| Fake-About | 17.4745 | <0.0001 | yes |
| Fake-Terms&Conditions | 108.1600 | <0.0001 | yes |

Table 3 shows the results of the logit model tested on our dataset. Model I includes all the variables, which coefficients are all significant at 1% except for the coefficient of the variable *About*. Model II yields similar results, however based on the Akaike information criterion we prefer it to model I. The signs of the coefficients are coherent with our expectations: if a website uses a security protocol, provides a "contact us" section, reports terms and conditions and a phone number, the probability that is a fake news website diminishes. The levels of VIFs for the variables is nearly 1, showing absence of multicollinearity, and the likelihood ratio test consents us to reject the null hypothesis that the model is not statistically significant. The level of the McFadden R-squared (0.4660) is moderate, however, this model is able to predict 337 case on 400 (Table 4), showing a prediction accuracy of the 84.2%. This result is satisfactory because an algorithm based on this model, with few predictors, is computationally fast and analyzes only the website to produce a probability of reliability.

Table 3. Results of the logit model based on (1). * indicates 1% significance.

|  | I | | | | II | | | |
|---|---|---|---|---|---|---|---|---|
|  | Coefficient | P-value | Slope | VIF | Coefficient | P-value | Slope | VIF |
| Constant | 3.7723* | <0.0001 | | | 3.8405* | <0.0001 | | |
| Padlock | −2.3133* | <0.0001 | −0.5058 | 1.392 | −2.3141* | <0.0001 | −0.5053 | 1.392 |
| Contact | −1.3385* | 0.0049 | −0.3064 | 1.606 | −1.1682* | 0.0089 | −0.2714 | 1.347 |
| Telephone | −1.7285* | <0.0001 | −0.4060 | 1.357 | −1.7179* | <0.0001 | −0.4040 | 1.356 |
| About | 0.3744 | 0.2789 | 0.0932 | 1.331 | | | | |
| Terms | −1.5144* | <0.0001 | −0.3605 | 1.402 | −1.4569* | <0.0001 | −0.3478 | 1.381 |
| McFadden R-squared | 0.4681 | | | | 0.4660 | | | |
| Adjusted R-squared | 0.4465 | | | | 0.4479 | | | |
| Akaike criterion | 306.9277 | | | | 306.1254 | | | |
| Likelihood ratio test chi-square (p-value in brackets) | 259.59 [0.0000] | | | | 258.392 [0.0000] | | | |
| Observations | 400 | | | | 400 | | | |

Table 4. Confusion matrix of model II.

|  | Predicted: Reliable | Predicted: Fake |
|---|---|---|
| Actual: Reliable | 166 (41.5%) | 34 (8.5%) |
| Actual: Fake | 29 (7.3%) | 171 (42.7%) |

## 5. Conclusions

In this work we have proposed a possible tool to distinguish a reliable source of news from a deceiving one. The parsimony of our model consents to reach a solution rapidly using only information potentially present in every website. Promoting a more reliable informative system is essential to:

1. reduce the social impact of fake news such as the sense of mistrust relatively to information diffused via classical and innovative means of communication;
2. reduce the economic impact of fake news such as prices far from the real value of an asset or the economic damage suffer by social networks and information companies;
3. reestablish trust in the institutions;
4. reward adequately (in terms of shares and notoriety) reputable news websites.

Future topics that we have not assessed in this paper are: how to enlarge the dataset in order to increase the number of predictors and the accuracy of the model; what sampling method could be adopted to maximize the power of the algorithm; what are the consequences of different levels of $T$, the threshold on which a user decides if a news should be shared or not. This last point is particularly interesting also from a theoretical economic perspective. For example, lower level of $T$ means that users are disposable to accept low-quality information. This may have profound implications triggering the undesired effect of "bad" information that replaces "good" information. So, our final question is: why should good information sources invest time and money in better and reliable articles if bad ones are shared by consumers without distinction? The answer will be content of future works.

**Funding**

This research did not receive any specific grant from funding agencies in the public, commercial, or not-for-profit sectors.